# Partial Redundancy Elimination for Multi-threaded Programs

Mohamed A. El-Zawawy[†] and Hamada A. Nayel[††],

[†]College of Computer and Information Sciences, Al-Imam M. I.-S. I. University, Riyadh, Kingdom of Saudi Arabia
[†]Department of Mathematics, Faculty of Science, Cairo University, Giza 12613, Egypt
[††]Faculty of Computers and Informatics, Benha University, Benha, Egypt

**Summary**
Multi-threaded programs have many applications which are widely used such as operating systems. Analyzing multi-threaded programs differs from sequential ones; the main feature is that many threads execute at the same time. The effect of all other running threads must be taken in account. Partial redundancy elimination is among the most powerful compiler optimizations: it performs loop-invariant code motion and common subexpression elimination. We present a type system with optimization component which performs partial redundancy elimination for multi-threaded programs.

## 1. Introduction

There are many methods for compiler optimizations; a powerful one of them is partial redundancy elimination (PRE). PRE eliminates redundant computations on some but not necessarily all paths of programs. PRE is a complex optimization as it consists of loop invariant code motion and common subexpression elimination. PRE was established by Morel and Renvoise [17] where they introduce a more general problem (as a system of Boolean equations). Xue and Cai formulated a speculative PRE as a maximum flow problem [27]. Xue and Knoop proved that the classic PRE is a maximum flow problem [28]. Saabas and Ustaluu use the type-systems framework to approach this problem [22]. Some optimizations have been added to PRE such as strength reduction [11] and global value numbering [3]. All methods mentioned above are established to operate on sequential programs.

In the present paper, we achieve partial redundancy elimination for multi-threaded programs which are widely used. Operating system is an example of a system software that depends on multi-threading. You can write your document in a word processor while running an audio file, downloading a file from the internet, and/or scanning for viruses (each of these tasks is considered a thread of computations). Web browser as an example can explore your e-mail, while downloading a file in the background. The key feature of multi-threaded programs is that many threads can be executed at the same time. Consequently, when executing a thread there is an effect that comes from executing other threads. In general, when analyzing multi-threaded programs, the effect of all threads at the same time must be taken in account. Hence, analyzing multi-threaded programs completely differs from sequential ones. Deducing and stating properties of programs can be done using type systems as well as program analysis. Program analysis has algorithmic manner while type systems are more declarative and easy to understand with type derivations that provide human-friendly format of justifications. We present a type system for optimizing multi-threaded programs. Our type system depends on a new analysis, namely *modified analysis,* and a function called *concurrent modified*, rather than on *anticipability* analysis and *conditional partial availability* analysis used for the while language.

Organization of this paper is as follow. In section 2 we introduce an operational semantics for the language we study. Section 3 presents the concepts of modified analysis and concurrent modified function. Also the soundness of modified analysis, the anticipability analysis, and conditional partial availability analysis for multi-threaded language are discussed in this section. In section 4, we present the type system including the optimization component and prove its soundness. Section 5 and 6 outline related and future work, respectively.

## 2. Motivation

In this section we introduce the language we study (*FWHILE*), a motivating example, and a natural semantics of *FWHILE*.

### 2.1 *FWHILE* Language

We assume that our reader is familiar with data flow analysis. We introduce a motivating example to show the importance and obstacles of applying PRE on multi-threaded programs. We use a simple language which we call *FWHILE*. The basic building blocks of *FWHILE* are literals $l \in \text{Lit}$, statements $s \in \text{Stmt}$, arithmetic expressions $a \in \text{AExp}$, and Boolean expressions $b \in \text{BExp}$. These blocks are defined over a set of program variables $x \in \mathbf{Var}$ and numerals $n \in \mathbf{Z}$ in the following way:

$$l ::= x \mid n$$
$$a ::= l \mid l_0 + l_1 \mid l_0 * l_1 \mid ...$$
$$b ::= l_0 = l_1 \mid l_0 \leq l_1 \mid ...$$
$$s ::= x := a \mid \mathbf{skip} \mid s_0; s_1 \mid \mathbf{if}\ b\ \mathbf{then}\ s_t\ \mathbf{else}\ s_f$$
$$\mid \mathbf{while}\ b\ \mathbf{do}\ s_t \mid \mathbf{fork}\{s_1, s_2, ..., s_n\}.$$

We use the following notation $\mathbf{AExp}^+$ for the non-trivial arithmetic expressions (i.e $\mathbf{AExp}^+ = \mathbf{AExp} \setminus \mathbf{Lit}$ ).

2.2 Motivating Example

The following is an example that motivates our research.

$$v := a - c; u := a + b;$$
$$\mathbf{fork}\{\{y := a+b; c := 2; z := a - c;\};$$
$$\{x := a+b; z := a - c;\};\}$$

In this example, expressions $a+b$ and $a-c$ are evaluated before reaching the fork statement, hence we can use their values in the fork statement. But, one of the threads modifies the value of $c$ hence we cannot use expressions containing $c$; because we do not know when the thread contains this modification will be executed. After applying our analysis the optimized version of the program will be:

$$v := a - c;$$
$$t_1 := a + b;$$
$$u := t_1;$$
$$\mathbf{fork}\ \{\{y := t_1; c := 2; z := a - c;\};$$
$$\{x := t_1; z := a - c;\}; \}$$

2.3 Natural Semantics

We use the semantics introuced by Mohamed El-Zawawy in [9]. We review this semantics in this section.
a *state* as a function from a set of variables to integers: $\sigma \in \mathrm{State}$, $\sigma: \mathrm{Var} \to \mathbf{Z}$. The state assigns a value for each variable. Expressions (arithmetic and boolean) are defined by semantic functions $[\![-]\!] \in \mathbf{AExp} \cup \mathbf{BExp} \to \mathbf{Z} \cup \{tt, ff\}$ in denotational style. For $a \in \mathbf{AExp}$ and $b \in \mathbf{BExp}$ we write $[\![a]\!]\sigma$ and $[\![b]\!]\sigma$ to denote the evaluations of expressions $a$ and $b$ in a state $\sigma$, respectively. We write $\sigma \models b$ to denote that $[\![b]\!] = tt$ (i.e evaluation of $b$ in $\sigma$ is true). Statements are written in the form of evaluation relation $\succ - - \to\ \subseteq \mathbf{State} \times \mathbf{Stm} \times \mathbf{State}$. The notation $\sigma[x \mapsto [\![b]\!]\sigma]$ denotes that the state is $\sigma$ rather than $\sigma(x) = [\![b]\!]\sigma$. Inference rules of the semantics are:

$$\dfrac{}{\sigma \succ x := a \to \sigma[x \mapsto [\![a]\!]\sigma]}\ :=_{ns} \qquad \dfrac{}{\sigma \succ \mathbf{skip} \to \sigma}\ skip_{ns}$$

$$\dfrac{\sigma \succ s_0 \to \sigma'',\ \sigma'' \succ s_1 \to \sigma'}{\sigma \succ s_0; s_1 \to \sigma'}\ comp_{ns} \qquad \dfrac{\sigma \models b,\ \sigma \succ s_t \to \sigma'}{\sigma \succ \mathbf{if}\ b\ \mathbf{then}\ s_t\ \mathbf{else}\ s_f \to \sigma'}\ if^{tt}_{ns}$$

$$\dfrac{\sigma \not\models b,\ \sigma \succ s_f \to \sigma'}{\sigma \succ \mathbf{if}\ b\ \mathbf{then}\ s_t\ \mathbf{else}\ s_f \to \sigma'}\ if^{ff}_{ns} \qquad \dfrac{\sigma \not\models b}{\sigma \succ \mathbf{while}\ b\ \mathbf{do}\ s_t \to \sigma}\ while^{ff}_{ns}$$

$$\dfrac{\sigma \models b,\ \sigma \succ s_t \to \sigma'',\ \sigma'' \succ \mathbf{while}\ b\ \mathbf{do}\ s_t \to \sigma'}{\sigma \succ \mathbf{while}\ b\ \mathbf{do}\ s_t \to \sigma'}\ while^{tt}_{ns}$$

$$\dfrac{\sigma_i \succ s_{\theta(i)} \to \sigma_{i+1},\ \forall i \in \{1,2,...\}}{\sigma_1 \succ \mathrm{fork}\{s_1, s_2, ..., s_n\} \to \sigma_{n+1}}\ fork_{ns}$$

where $\theta$ is a permutation on $\{1, 2, ..., n\}$

The rule $fork_{ns}$ depends on the the order of executing the threads. We assume that order named $\theta$ which is a permutation on $n$ (number of threads). In this rule we assume that the threads will execute one by one.

## 3. Program Analysis

In this section we will introduce the analysis of the multithreaded programs. We introduce type systems to help optimizing programs. Firstly, we introduce the modified analysis which tells which variables are modified. Secondly, we introduce a concurrent modified function. Also, we introduce traditional anticipability analysis and conditional partial availability analysis which are generalizations of the work of [22] (with additional rules for multi-threaded statements).

3.1 Modified Analysis

Modified analysis computes for each program point which variables have been modified. The type system is simple. It gathers the modified variables along the path to the point. Type $m \subseteq \mathbf{Var}$ is a set of variables. Modified analysis is a must forward analysis. The subtyping is the revered set inclusion (i.e $\leq\ =\ \supseteq$).

**Definition 1** *For any program point, any state $\sigma$ and a type $m \subseteq \mathbf{Var}$, e $\sigma \models m$ ( $\sigma$ entails $m$ ) iff $m \subseteq dom(\sigma)$.*

Type system is as follow :

$$\frac{}{x := a : m \to m \cup \{x\}} \; m :=$$

$$\frac{s_0 : m \to m'', \; s_1 : m'' \to m'}{s_0 ; s_1 : m \to m'} \; mcomp$$

$$\frac{s_t : m \to m}{\textbf{while } b \textbf{ do } s_t : m \to m} \; mwh$$

$$\frac{}{\textbf{skip} : m \to m} \; mskip$$

$$\frac{s_t : m \to m' \quad s_f : m \to m'}{\textbf{if } b \textbf{ then } s_t \textbf{ else } s_f : m \to m'} \; mif$$

$$\frac{m \le m_0 \quad s : m_0 \to m_0' \quad m_0' \le m'}{s : m \to m'} \; mconseq$$

$$\frac{s_{\theta(i)} : m \to m_i \quad \forall i \in \{1,2,\ldots n\}}{\textbf{fork}\{s_1, s_2, \ldots, s_n\} : m \to \bigcup_{1 \le i \le n} m_i} \; mfork$$

The type system is clear and simple. The rule $m :=$ adds the assigned variable to pre type. Rules $mskip$, $mconseq$, $mif$ and $mwh$ are direct and similar to operational semantics. Rule $mconseq$ for strengthen the pre-type. Rule $mfork$ is for threading. This rule computes the modified variables along **fork** statement by collecting all modified variables over all threads.

The following two lemmas state properties about modified analysis:-

**Lemma 1** *Suppose* $s : m \to m'$, *where* $m, m' \subseteq \textbf{Var}$. *Then* $m \subseteq m'$.

*Proof :* It is clear that the statement which actually changes the set $m$ is the assignment statement where $m' = m \cup \{x\}$ i.e $m \subseteq m'$.

**Lemma 2** *Suppose* $s : m' \to m''$ *and* $\sigma \succ s \to \sigma'$, *where* $m, m', m'' \subseteq \textbf{Var}$. *Then* $s : m' \cup m \to m'' \cup m$.

*Proof :* We have $m \subseteq dom(\sigma) \Rightarrow m' \subseteq dom(\sigma')$. Then it is clear that :-
$$m \cup m'' \subseteq dom(\sigma) \Rightarrow m' \cup m'' \subseteq dom(\sigma')$$

### 3.2 Soundness of modified analysis

The following theorem proves the soundness of modified analysis.

**Theorem 1**
*Suppose* $s : m \to m'$ *and* $\sigma \succ s \to \sigma'$. *Then*
*if* $\sigma \models m$ *then* $\sigma' \models m'$.

*Proof :* The proof is by structure induction of type derivation. We will prove only main rules.

• Type derivation is $m :=$ and corresponding operational semantic is $:=_{ns}$:

We have $\sigma \models m$ and $\sigma' = \sigma[x \mapsto [\![a]\!]\sigma]$, which implies
$dom(\sigma') = dom(\sigma) \cup \{x\}$.
$\because m \subseteq dom(\sigma) \Rightarrow m \cup \{x\} \subseteq dom(\sigma) \cup \{x\}$
$\Rightarrow m' \subseteq dom(\sigma') \Rightarrow \sigma' \models m'$

• Type derivation is $mconseq$ and $\sigma \succ s \to \sigma'$:

$\sigma \models m \Rightarrow \sigma \models m_0 \quad (m \le m_0)$
$\Rightarrow \sigma' \models m_0' \quad (s : m_0 \to m_0' \; \sigma \succ s \to \sigma')$
$\Rightarrow \sigma' \models m' \quad (m_0' \le m')$

i.e $\quad \sigma \models m \Rightarrow \sigma' \models m'$

• Type derivation is $mfork$ and the corresponding operational semantic is $fork_{ns}$. We prove that :
$$\sigma_1 \models m_1 \Rightarrow \sigma_{n+1} \models \bigcup_{1 \le i \le n} m_i$$

From premises we have $s_{\theta(i)} : m \to m_i$, which by lemma 1 implies $m \subseteq m_i, \forall i \in \{1,2\ldots,n\}$.

From lemma 2 and lemma 1 we can get:
$s_{\theta(1)} : m \to m_1$
$s_{\theta(2)} : m_1 \to m_1 \cup m_2$
$s_{\theta(3)} : m_1 \cup m_2 \to m_1 \cup m_2 \cup m_3$
$\vdots$
$s_{\theta(n)} : m_1 \cup m_2 \ldots \cup m_{n-1} \to m_1 \cup m_2 \ldots \cup m_n$

To simplify the notations we let
$M_0 = m$,
$M_1 = m_1$,
$M_2 = m_1 \cup m_2$
$\vdots$
$M_n = m_1 \cup m_2 \ldots \cup m_n = \bigcup_{1 \le i \le n} m_i$.

The previous sequence can be written as:-
$s_{\theta(i)} : M_{i-1} \to M_i$, also we have $\sigma_i \succ s_{\theta(i)} \to \sigma_{i+1}$.
Then we get:
$\sigma_1 \models M_0 \Rightarrow \sigma_2 \models M_1$
$\sigma_2 \models M_1 \Rightarrow \sigma_3 \models M_2$
$\vdots$
$\sigma_n \models M_{n-1} \Rightarrow \sigma_{n+1} \models M_n$

From last sequence, we conclude $\sigma_1 \models m \Rightarrow \sigma_{n+1} \models M_n$.

## 3.3 Concurrent modified function $C$

In this section we will present concurrent modified function $C$. We start by defining *sub-statement* relation between statements.

**Definition 2** *For any two statements $s$ and $t$, we say that $t$ is a sub-statement of $s$ written as* $(t \prec s)$ *iff*:-

1. $t = s$ or

2. $s = s_1; s_2$ and $(t \prec s_1)$ or $(t \prec s_2)$

We mean by $t = s$, that $t$ is identically (syntactically) equivalent to $s$.

**Definition 3** *The modified concurrent function assigns a set of variables for each program point,* i.e. $C : \mathbf{Stm} \to \mathbf{Var}$:

1. For $\mathbf{fork}\{s_1, \ldots, s_n\}$ statement:

$$C(s_i) = \bigcup_{i \neq j} m_j, \quad \text{where} \quad s_i : m \to m_i$$

$$C(t) = C(s_i) \quad \text{for} \quad t \prec s_i$$

2. otherwise

$$C(s) = \varphi$$

## 3.4 Anticipability analysis

We now present anticipability analysis. For each program point it computes which non-trivial arithmetic expressions will be evaluated on all paths before any of their operands are modified. In the typing rule we use $eval(a)$ to denote $\{a\}$ if $a$ is non-trivial expression and $\varphi$ otherwise.

For $a \in \mathbf{AExp}^+$ we define :-

$mod(x) =_{df} \{a \mid x \in FV(a)\}$ and

$mce(s) =_{df} \{a \mid a \in mod(y). \forall y \in C(s)\}$

We use $\underline{s}$ to denote the full type derivation of $s : m \to m'$. Inference rules of the type system include:

$$\overline{\underline{x := a} : (ant' \setminus mod(x) \cup eval(a)) \setminus mce(x := a) \to ant'}$$

$$\frac{\underline{s_i} : ant_i \to Ant' \setminus mce(s_i) \quad \forall i \in \{1,2,\ldots n\}}{\mathbf{fork}\{\underline{s_1}, \underline{s_2}, \ldots \underline{s_n}\} : \bigcup_{1 \leq i \leq n} ant_i \to Ant'}$$

The rules for other program statements (if statement, while statement,..) follow the same line of corresponding rules introduced in [22] for the while langauge. The novelty of our work comes from fitting the fork statement into the type system of [22] and making necessary changes. We note that if no thread exists then for each statement $s$ in the program $mce(s) = \varphi$. Besides the rule of **fork** statement which characterizes the multi-threaded concept. These rules prevent any modified expression (i.e modified in concurrent threads) from being used from the start of **fork** statement. As anticipability analysis is backward analysis, we remove modified expressions from $Ant'$ of each statement. Also, in each assignment statement we remove modified expressions. All of these removals are guided by the set $mce(s)$.

## 3.5 Partial availability analysis

It computes for each program point which non-trivial arithmetic expressions has already been evaluated and later not modified on some path through this program point and also anticipable.
Inference rules of the type system include:

$$\frac{\underline{s_i} : ant_i, CPAV \setminus mce(s_i) \to Ant' \setminus mce(s_i), cpav'_i \quad \forall i \in \{1,2,\ldots,n\}}{\mathbf{fork}\{\underline{s_1}, \underline{s_2}, \ldots, \underline{s_n}\} : \bigcup_{1 \leq i \leq n} ant_i, CPAV \to Ant', \bigcup_{1 \leq i \leq n} cpav'_i}$$

The rules for other program statements (if statement, while statement,..) follow the same line of corresponding rules introduced in [22] for the while langauge. The novelty of our work comes from fitting the fork statement into the type system of [22]. Here we excluded the expressions in concurrent modified of all threads of **fork** statement to avoid using these expressions after exiting **fork** statement.

## 4. Optimization Component

In this section we introduce a type system with optimization components for multithreaded programs. We mean by the notation $\underline{s} : ant, cpav \to ant', cpav' \triangleright s_*$ that, the statement $s$ with complete type system $m, ant$ and $cpav$ is optimized to $s_*$. Inference rules of the type system include:

$$\frac{\underline{s_i} : ant_i, CPAV \setminus mce(s_i) \to Ant' \setminus mce(s_i), cpav'_i \triangleright s^* \quad \forall i \in \{1,2,\ldots,n\}}{\mathbf{fork}\{\underline{s_1}, \underline{s_2}, \ldots, \underline{s_n}\} : \bigcup_{1 \leq i \leq n} ant_i, CPAV \to Ant', \bigcup_{1 \leq i \leq n} cpav'_i \triangleright \mathbf{fork}\{s_1^*, s_2^*, \ldots s_n^*\}} fork_{pre}$$

The rules for other program statements (if statement, while statement,..) follow the same line of corresponding rules introduced in [22] for the while langauge. The novelty of our work comes from fitting the fork statement into the type system of [22]. Our new analysis also affects the component of optimizations as in the rule **fork**$_{pre}$, which achieves optimizion via optimizing each thread of its components.

The soundness of the above system is proved in [22] except of course for the fork rule which is at the heart of our contribution. We discuss this case here. The definition of the similarity relation $\sigma \sim_{cpav} \sigma_*$ is in [22]. Soundness of the fork statement is as follows:

Suppose $s: ant, cpav \rightarrow ant', cpav', \sigma \sim_{cpav} \sigma_*$ and $\sigma \succ s \rightarrow \sigma'$. We have to find $\sigma_*'$ such that $\sigma' \sim_{cpav'} \sigma_*'$ and $\sigma_* \succ s_* \rightarrow \sigma_*'$. The proof is by induction on the typing derivation. We have $s_i : ant_i, cpav \setminus mce(s_i) \rightarrow ant' \setminus mce(s_i), cpav_i' \triangleright s_i^*$.

If **fork** $\{\underline{s}_1, \underline{s}_2, ..., \underline{s}_n\} : a, c \rightarrow a', c' \triangleright$ **fork** $\{s_1^*, s_2^*, ..., s_n^*\}$ and $\sigma_1 \sim_c \sigma_1^*$, where $a = \bigcup_{1 \leq i \leq n} ant_i, c = cpav, a' = Ant', c' = \bigcup_{1 \leq i \leq n} cpav_i$ then we have to find $\beta$ such that:

if $\sigma_1 \succ \textbf{fork}\{s_1, s_2, ..., s_n\} \rightarrow \sigma_{n+1}$ then $\sigma_{n+1} \sim_{c'} \beta$ and $\sigma_i^* \succ \textbf{fork}\{s_1^*, s_2^*, ..., s_n^*\} \rightarrow \beta$. The following is given: if $s_i : ant_i, cpav \setminus mce(s_i) \rightarrow ant' \setminus mce(s_i), cpav_i' \triangleright s_i^*$ and $\sigma_j \sim_{c_i} \sigma_j^*$ (where $c_i = cpav \setminus mce(s_i)$) and $j = \theta(i)$) then $\sigma_i \succ s_j \rightarrow \sigma_{i+1}$ implies the existence of $\gamma_{i+1}^*$ such that $\gamma_{i+1}^* \sim_{cpav_j} \sigma_{i+1}$ and $\sigma_i^* \succ s_j \rightarrow \gamma_{i+1}^*$.

Choosing $\gamma_{i+1}^* = \sigma_{i+1}$, our choice is amenable where, it leads to the following:

$\sigma_i^* \succ s_j^* \rightarrow \sigma_{i+1}^* \Rightarrow \sigma_1^* \succ \textbf{fork}\{s_1^*, s_2^*, ..., s_n^*\} \rightarrow \sigma_{n+1}^* \cdots i$

$\sigma_{i+1} \sim_{cpav_j} \sigma_{i+1}^* \Rightarrow \sigma_{n+1}^* \sim_{c'} \sigma_{n+1} \quad\quad \cdots ii$

from $i$ and $ii$ and choosing $\beta = \sigma_{n+1}^*$ the proof is done.

## 5. Related Work

There are three aspects of work that are related to our work:-

**Partial Redundanc elimination (PRE)**
PRE [8,18,25] was originated by Morel and Renvoice[17]. They applied PRE using static analysis and presented PRE as a general problem of global optimization using boolean system of equations. [17] also presents an algorithm for global optimization, which does not need *a control flow graph* (CFG). Efforts have been done to improve the formulation of PRE [7]. The work in [28,7] formulates classical and commutative PRE as a maximum flow problem. PRE is used as a framework and is extended to do more optimizations as strength reduction [11]

*Type systems*
Analyzing a program using type systems rather than *control flow graph* was the idea of work in [14]. More accurately, it proves that types can describe results of an analysis of a program if and only if this type is a supertype of a result of applying the analysis. The work in [1,9,10,14,18,19,21,22] uses type systems to accomplish the program analysis in similar way of ours. The work in [21] introduces type systems for PRE and proves the correctness of optimizations. Dead code elimination was treated using type systems. [1] presents type systems for dead code elimination and constant folding. Also [22] introduces type systems for dead code elimination and common subexpression elimination and proves a relational soundness.

*Analysis of Multi-threaded Programs*
The field of program analysis has been extended to treat multi-threaded programs besides sequential ones. We can use these analyses in compiler optimizations. The main obstacle to apply traditional compiler optimizations is that, we do not know which order of threads will be executed [15]. Generalizing standard program presentations, analysis, and transformations are used to optimize multi-threaded programs in the presence of access to shared data [12,13,24,26]. Another way to use analysis occurs when two threads access the same data without synchronization (one of them is write): data races[5,6,16]. Yet another aspect to use analysis is called dead lock detection (which occurs when threads are permanently blocked waiting for resources)[2,4]. Pointer analysis [9,20,23] of multi-threaded programs attracts attention of researchers. None of works above deal with PRE. The present paper is the first work that uses type systems as a framework to implement PRE to multi-threaded programs.

## 6 Conclusions and Future Work

In this paper the main contribution is the application of PRE to a multi-threaded programming language. Up to our knowledge, this paper is the first to deal with this problem. We use type systems as a tool to solve the problem. We designed a simple type system for optimizing multi-threaded programs. We approach the problem in a simple way; we use usual PRE with simple modifications. We look for variables that have been modified in other threads and exclude the expressions that contain any of the modified variables. For future work, we study more complicated optimization and consider using other tools. Many modifications can be applied.

Zeller, editors, *Compiler Construction*, volume 3923 of *Lecture Notes in Computer Science*, pages 139-154.Springer Berlin / Heidelberg, 2006.

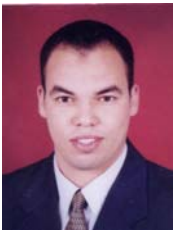

**Mohamed A. El-Zawawy** received: PhD in Computer Science from the University of Birmingham in 2007, M.Sc. in Computational Sciences in 2002 from Cairo University, and a BSc. in Computer Science in 1999 from Cairo University. Dr El-Zawawy is an assistant professor of Computer Science at Faculty of Science, Cairo University Since 2007. Currently, Dr. El-Zawawy is on a sabbatical from Cairo University to College of Computer and Information Sciences, Al-Imam M. I.-S. I. University, Riyadh, Kingdom of Saudi Arabia. During the year 2009, Dr. El-Zawawy held the position of an extra-ordinary senior research at the Institute of Cybernetics, Tallinn University of Technology, Estonia. Dr. El-Zawawy worked as a teaching assistant at Cairo University from 1999 to 2003 and latter at Birmingham University from 2003 to 2007. Dr. El-Zawawy is interested in static analysis, shape analysis, type systems, and semantics of programming languages.

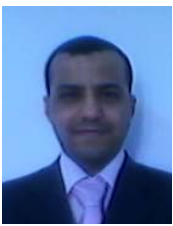

**Hamada A. Nayel** received the B.Sc. in mathematics from Benha University in 2003. Diploma in computer science and information systems from Institute of Statistical Studies and Research (ISSR), Cairo University in 2005. He worked as a demonstrator in faculty of Science from 2003 to 2009. Since 2009, he works at faculty of Computers and Informatics, Benha University. Mr. Nayel is interested in semantics of programming languages, type systems, static analysis, and shape analysis.